\documentstyle[preprint,revtex]{aps}
\begin{document}

\draft

\begin{title}
Relation between the phenomenological interactions of the
algebraic cluster model and the effective two--nucleon forces
\end{title}

\author{K. Varga and  J. Cseh}

\begin{instit}
Institute  of Nuclear Research of the Hungarian Academy of
Sciences (MTA ATOMKI) Debrecen, Hungary
\end{instit}

\begin{abstract}
We determine the phenomenological cluster--cluster interactions
of the algebraic model corresponding to the most often used
effective two--nucleon forces for the $^{16}$O + $\alpha$
system.
\end{abstract}


\narrowtext

\section{Introduction}

More than ten years ago the vibron model was proposed as a
phenomenological algebraic description of the nuclear cluster or
molecular states
\cite{ia81}.
This is a model of the dipole collective motion, which is
applied also in molecular
\cite{il82}
and in hadron spectroscopy
\cite{ia89}.
Its formalism is very similar to that of the Interacting Boson
Model of the quadrupole collectivity of nuclei
\cite{ia87}.

The interaction of the clusters in the vibron model is treated
in a phenomenologic way, i.e.  the Hamiltonian is expressed in
terms of boson operators, and the expansion coefficients are
fitted to experimental data. It is an interesting question, how
these phenomenological algebraic cluster--cluster interactions are
related to the effective two--nucleon forces, which are applied
e.g. in microscopic cluster studies
\cite{hi86}.
Except for a brief discussion in
\cite{cs91t}
this question has not been investigated so far, although it is of
great importance from the viewpoint of the
microscopic foundation of the algebraic cluster model. The
present paper is meant to be a contribution to this task.

The application to some well--known cluster bands in light
nuclei revealed that i) the $U(4) \supset U(3) \supset O(3)$
basis of the vibron model is preferred to the other possible
basis, and ii) this basis has to be truncated in a well--defined
way in order to get rid of the Pauli forbidden states
\cite{cl88,cs89,cs91}.
In particular, the $n_{\pi}$ quantum number which is the
representation label of the $U(3)$ group and gives the number of
oscillator quanta in the relative motion of the two clusters,
has to be larger than a limit obtained from the Wildermuth condition
\cite{wt77}.
By taking into account the Pauli blocking in this way, the model
space of the algebraic description becomes a subset of the model
space of the microscopic $SU(3)$ cluster model
\cite{hi86,wt77}.
(Due to the finite value of the $U(4)$ representation index
$n_{\pi}$ has an upper limit, too.)

When the internal degrees of freedom of the clusters play an
important role, i.e. in case of non--closed--shell clusters, the
model space is larger and the group structure of the algebraic
description is more complicated. Nevertheless, for systems of
non--closed--shell clusters the model space can also be
constructed in such a way that it is free from the Pauli
forbidden states and the spurious center of mass motion
\cite{cs91t,cs92l}.
It is done by using the $SU(3)$ shell model
for the description of the internal cluster degrees of freedom,
instead of the previously applied phenomenological interacting
boson
\cite{di86},
or interacting fermion
\cite{lc91}
models. Again, the model space is a subspace of that of the
microscopic $SU(3)$ cluster model.

The algebraic approach, in which the model space is free from
the forbidden states and the interactions are treated
phenomenologically is called semimicroscopic
algebraic description. Since the basis states of this
description have a one--to--one correspondence with the $SU(3)$
basis states of the microscopic cluster models, one can relate
the phenomenological cluster--cluster interactions to the
effective two--nucleon forces simply by equating the
corresponding matrix elements. Here we present such a relation
for the example of the
$^{16}O + \alpha$ system.

In what follows, in Section II. we give
the matrix elements of the vibron model Hamiltonian, and show
how it fits the energy spectrum of some selected bands in the
$^{20}$Ne nucleus.
In Section III. the matrix
elements of some frequently applied effective two--nucleon
forces  are calculated.
Finally, the relation between the phenomenological and
microscopic interactions are discussed in Section IV.

\section{The phenomenological Hamiltonian and its matrix elements}

In the algebraic description of clusterization the spectrum is
generated by the interactions of a finite number ($N$) of
bosons, which can occupy single--particle states with angular
momentum and parity: $0^+$ ($\sigma$ bosons) and $1^-$ ($\pi$
bosons). The total number of particles is conserved, therefore
the creation and annihilation operators appear only in number
conserving bilinear forms. They generate the $U(4)$ group, and
the group structure of the model manifests itself in a twofold
way: not only the basis states are characterized by the
representation labels of the group--chain
\[
U(4) \supset U(3) \supset O(3), \label{mlett:1}
\]
\begin{equation}
\vert\ \  N \ \ , \ \ \ n_{\pi} \ \ , \ \ L \ \ \rangle .
\label{1}
\end{equation}
\[
n_{\pi} = N, N-1,...,0; \ \ L = n_{\pi}, n_{\pi}-2,...,1 \ or \ 0,
\]
but also the physical operators are obtained in terms of the
generators of the $U(4)$ group
\cite{il82}.
(We consider here, as mentioned in the Introduction, the simple
case of closed--shell clusters.) An especially important
limiting situation, called dynamical symmetry, is reached, when
the Hamiltonian can be expressed in terms of the Casimir invariants
of the group--chain
(\ref{1}).
Then the eigenvalue problem has an analytical solution. If we
consider only one-- and two--body interactions, the Hamiltonian
of the $U(3)$ dynamical symmetry can be written as:
\begin{equation}
H = h_0 C_{1U4} + h_1 C_{2U4} + h_2 C_{1U3} + h_3 C_{2U3} +
h_4 C_{2O3} \ ,
\label{2}
\end{equation}
where the $C$'s stand for the Casimir operators of the indicated
order, e.g. $C_{2U3}$ is the second order Casimir of the $U(3)$
group, and $h_i$'s are phenomenological parameters. The
Hamiltonian matrix is diagonal in the basis
(\ref{1}),
and the energy eigenvalues are:
\begin{equation}
E = \epsilon + \beta L(L+1) + \gamma n_{\pi} + \delta n_{\pi}^2  .
\label{3}
\end{equation}
Here $\beta = h_3$, $\gamma = h_2 +3h_3$, $\delta = h_3$,
and the eigenvalues of the
$U(4)$ Casimir operators could be involved in the constant
$\epsilon$, because $N$ is conserved.

Our aim here is to relate these phenomenological parameters to
the effective two--nucleon forces. Nevertheless, in order to
illustrate the ability of the vibron model Hamiltonian
containing only one-- and two--body terms, it is worthwhile to
show to what extent can it fit an experimental spectrum. For
this purpose we have chosen the $^{20}$Ne nucleus, because its
$K^{\pi} = 0_1^+, \ 0^-,\ 0^+_4$
bands are
known to have a well--developed $^{16}O + \alpha$ cluster
structure
\cite{fu80}.
In addition, more recently the
$K^{\pi} = 0_5^+$
band has been established based on alpha--scattering data
\cite{ri84},
although the assignments of the $J \ge 4$ spins are less
certain. The experimental spectrum that we have considered
\cite{ri84,aj87}
is plotted on the left hand side of Fig. 1. Their description
in terms of the Hamiltonian
(\ref{2},\ref{3})
with $U(3)$ dynamical symmetry is shown on the right hand side.
For the $^{16}O + \alpha$ system the Wildermuth condition gives
$n_{\pi} \ge 8$
\cite{cl88,cs89}.
The parameters of the energy--expression
(\ref{3})
were obtained from a least--square fitting procedure, in which
the weight of the uncertain states were 0.5, in comparison with
the usual weight of 1.0 of the fully established states. (For
the $6^+$ and $8^+$ members of the $0^+_5$ band we have chosen
the total weight of the 2 or 3 candidates to be 0.5.) The
parameter values are (in $MeV$):
$\beta = 0.161, \
\gamma = 13.601, \
\delta = -0.571, \
\epsilon = -71.040.$

\section{Matrix elements of the effective two--nucleon interactions}

In the microscopic cluster model
the wave function of the $^{20}$Ne
nucleus is given by
\begin{equation}
\Psi_{^{20}Ne}=n_0 {\cal A} \lbrace \phi_{\alpha} \phi_{^{16}O}
\chi({\bf r}_{\alpha-^{16}O})\rbrace
\label{4}
\end{equation}
where ${\cal A}$ is an intercluster antisymmetrizer
\cite{hori}
and
$\phi_{\alpha}$ and $\phi_{^{16}O}$ are normalized antisymmetric
internal wave functions of the alpha particle and the $^{16}$O
nucleus, respectively.
Furthermore, $\chi({\bf r}_{\alpha-^{16}O})$ is the wave
function of the relative motion and $n_0$ is a normalization constant.
If the internal states are harmonic oscillator
shell model ground states of common
size parameter $\nu (={m \omega \over 2 \hbar})$ (SU(3) scalars:
$(\lambda,\mu)=(0,0)$ ) and the wave function of the relative motion is
a harmonic oscillator wave function $\varphi_{nlm}$ of size parameter
${16 \times 4 \over 16 + 4 } \nu$ (belonging to
the $(2n+l,0)$ representation), then using the language of Elliott's SU(3)
group
\cite{elli}
\begin{equation}
\psi_{nl}=n_0 {\cal A} \lbrace \phi_{\alpha} \phi_{^{16}O} \varphi_{nlm}
({\bf r}_{\alpha-^{16}O}) \rbrace
\label{5}
\end{equation}
belongs to the SU(3) irreducible representation
$(2n+l,0)$
\cite{wk58,bb58}.
(We note here the $n_{\pi} = 2n + l$ relation of the quantum numbers
\cite{cl88,cs89}.)

The conventional technique to calculate the matrix elements of the microscopic
Hamiltonian
\begin{equation}
H=\sum_i T_i+\sum_{ij} V_{ij}
\label{6}
\end{equation}
(sandwiched between the wave function  $\psi_{nl}$) is based
on the fact that the
"shifted" gaussian function
\begin{equation}
\varphi_{\bf s}^{\alpha}({\bf r})=
\left({2 \alpha \over \pi}\right)^{3/4} e^{-\alpha({\bf r}-{\bf s})^2}
\label{7}
\end{equation}
is a generating function of the harmonic oscillator function
\cite {hori}:
\begin{equation}
\varphi_{nlm}({\bf r})=A_{nl} \int d{\hat s} Y_{lm}({\hat s})
{d^{2n+l} \over ds^{2n+l}}  \varphi^{\alpha}_{\bf s} ({\bf r})
e^{{\alpha \over 2} {\bf s}^2} \left\vert _{s=0} \right. \ ,
\label{8}
\end{equation}
where
\begin{equation}
A_{nl}= (-1)^n \sqrt{{(2n+l)! \over 4\pi (2n)!!(2n+2l+1)!!}}.
\label{9}
\end{equation}
Using the equation
(\ref{8})
the matrix elements of the states
$\psi_{nl}$ can be derived as
\begin{equation}
\langle \psi_{nl} \vert H \vert \psi_{nl} \rangle=
A_{nl}^2 \int d {\hat s} Y_{lm}({\hat s}) \int d {\hat s'}
Y_{lm}({\hat s'})
{\partial^{4n+2l} \over \partial s^{2n+l} \partial {s'}^{2n+l} }
H({\bf s},{\bf s'}) \left\vert _{s=0 \atop s'=0} \right. \ \ \ .
\label{10}
\end{equation}
The matrix elements on the right hand side, the so called
generator coordinate method (GCM)
matrix elements of $^{20}$Ne (see e.g.
\cite{hori})
\begin{equation}
H({\bf s},{\bf s'})=
\langle {\cal A} \lbrace \phi_{\alpha} \phi_{^{16}O}
\varphi^{{16 \times 4 \over 16+4 } \nu}
_{\bf s} ({\bf r}_{\alpha-^{16}O}) \rbrace
\vert H \vert
{\cal A} \lbrace \phi_{\alpha} \phi_{^{16}O} \varphi^{{16 \times 4 \over
16+4 } \nu}_{\bf s'} ({\bf r}_{\alpha-^{16}O}) \rbrace
\rangle
\label{11}
\end{equation}
can be calculated easily. Owing to the well-known theorem of
Elliott and Skyrme
\cite{elsk},
one can express  the wave function
${\cal A} \lbrace \phi_{\alpha} \phi_{^{16}O} \varphi^{{16 \times 4 \over
16+4 } \nu}_{\bf s'} ({\bf r}_{\alpha-^{16}O}) \rbrace$
by a Slater determinant of harmonic oscillator single particle orbits
centered around ${\bf s}_{\alpha}$ and ${\bf s}_{^{16}O}$
\begin{eqnarray}
&& {\cal A} \lbrace \phi_{\alpha} \phi_{^{16}O} \varphi^{{16 \times 4 \over
16+4 } \nu}_{\bf s'} ({\bf r}_{\alpha-^{16}O}) \rbrace
=
\left({\pi \over 2\cdot 20 \nu }\right) ^{3/4}
e^{+20\nu({\bf S}-{\bf R})^2} \nonumber\\
&& \cdot \ {1 \over \sqrt{20!}}
det\lbrace
(000)_{{\bf s}_{^{16}O}}^{4}
(010)_{{\bf s}_{^{16}O}}^{4}
(011)_{{\bf s}_{^{16}O}}^{4}
(01-1)_{{\bf s}_{^{16}O}}^{4}
(000)_{\alpha}^4 \rbrace.
\label{12}
\end{eqnarray}
In this equation $(nlm)^{4}_{\bf s}$ stands for the harmonic oscillator
shell model orbit $nlm$ centered around ${\bf s}$
and filled by
four nucleons with different spin-isospin configuration.
For example
\begin{eqnarray}
(000)_{{\bf s}_{^{16}O}}^{4}=
&& \varphi_{000}({\bf x}_1-{{\bf s}_{^{16}O}}) \eta_{p\uparrow}(1)
\varphi_{000}({\bf x}_2-{{\bf s}_{^{16}O}}) \eta_{p\downarrow}(2)
 \nonumber\\
&& \cdot \
\varphi_{000}({\bf x}_3-{{\bf s}_{^{16}O}}) \eta_{n\uparrow}(3)
\varphi_{000}({\bf x}_4-{{\bf s}_{^{16}O}}) \eta_{n\downarrow}(4),
\label{13}
\end{eqnarray}
where ${\bf x}_i$ is the single particle coordinate,
$\eta_{\sigma \tau}(i)$ is the spin-isospin function
of the $i$th nucleon.
The parameter coordinates can be related as
${\bf s}={\bf s}_{^{16}O}-{\bf s}_{\alpha}$ and
${\bf S}={1 \over 20} (16{\bf s}_{^{16}O}+4{\bf s}_{\alpha})$,
and ${\bf R}$  is the center of mass  coordinate of the $^{20}$Ne.

Using the reduction formulae given by
\cite{brin},
the matrix elements between Slater
determinants can be expressed by the single particle matrix elements of
the Hamiltonian and by the overlap of the (nonorthogonal) single particle
orbits. To facilitate the analytical calculation the nucleon-nucleon
interactions  used in the microscopic cluster model are linear combination
of gaussian potentials
\begin{equation}
V_{ij}=\sum_{k} (W_{k}+M_{k}P_{ij}^{x}+B_{k}P_{ij}^{\sigma}+
H_{k}P_{ij}^{x}P_{ij}^{\sigma})V_k e^{-({\bf x}_i-{\bf x}_j)^2/d_{k}^2} ,
\label{14}
\end{equation}
where $P_{ij}^{x}$ is the spatial $P_{ij}^{\sigma}$ the spin exchange operator
between particles $i$ and $j$, $W_k$, $M_k$, $B_k$ and $H_k$ are the
Wigner, Majorana, Bartlett and Heisenberg parameters, $V_k$ is the strength
$d_k$ is the diffusity of the potential. Using this gaussian interactions
the matrix element $H({\bf s},{\bf s'})$ has the following form
\begin{equation}
H({\bf s},{\bf s'})=\sum_{i} c_i s^{2k_i} s'^{2k'_i} {\bf ss'}^{m_i}
e^{-a_i s^2-a'_is'^2+b_i{\bf ss'}}  .
\label{15}
\end{equation}
The $c_i$, $a_i$, $a'_i$ and $b_i$ constants are expressed by the
$W_k$, $M_k$, $B_k$, $H_k$, $V_k$ and $a_k$ parameters of the
potential and the $\nu$ parameter of the harmonic oscillator
wave function. The explicit analytical expressions are too lengthy to be
tabulated here and the interested reader can find them
in the references
\cite{tosu,aoho,maka}.

The normalization constant $n_0$ can be
determined in the same manner, putting the unity operator in
place of $H$.

In determining the SU(3) matrix elements numerically,
we have considered some of the most conventional effective
interactions used in the cluster model calculations:
a phenomenological one proposed by Volkov (Volkov force number 2 (V2))
\cite{V2},and a potential of reaction matrix type given
by Hasegawa and Nagata  (HN1 and HN2)
\cite{HN},
and a modified version of the latter force
\cite{MHN}.
All parameters of the forces NH1, NH2 and MHN are fixed. The
Volkov force has a parameter (Majorana exchange parameter)
for the strength of the odd-parity state. This parameter can be
adjusted to the separation energy of the clusters.
These forces had been successfully applied in GCM calculations
for $^{20}$Ne
\cite{makafu}
and for other light nuclei. The parameters of these effective
nucleon-nucleon interactions can be found in Table I.
As an illustrative
numerical example the matrix elements
$\langle \psi_{nl} \vert H \vert \psi_{nl} \rangle$
using V2 force are listed in Table II. In agreement with Matsuse
et. al. \cite{makafu} the harmonic oscillator size parameter and the Majorana
parameter of the Volkov force is
chosen to be
$\nu=0.16 fm^{-2}$, and $M_k=0.62 (k=1,2)$ respectively.

\section{Phenomenological interactions from microscopic forces}

The parameters of the phenomenological interactions
(\ref{2},\ref{3}) can be obtained from the effective
two--nucleon forces by equating the corresponding
matrix elements of the two descriptions.
Taking e.g. the
$H_{32}, \ H_{40}, \ H_{50}, \ H_{60}$
matrix elements, where we have used the simplified notation of
$H_{nl} = \langle \psi_{nl} \vert H \vert \psi_{nl} \rangle$,
with the straightforward relations:
\begin{eqnarray}
&& \beta = {1 \over 6} (H_{32} - H_{40}) \ ,
\nonumber \\
&& \gamma = {1 \over 4} (-11H_{40} + 20H_{50} - 9H_{60}) \ ,
\nonumber \\
&& \delta = {1 \over 8} (H_{40} - 2H_{50} + H_{60}) \ ,
\nonumber \\
&& \epsilon = (15H_{40} - 24H_{50} + 10H_{60}) \ ,
\label{16}
\end{eqnarray}
and using their analytical expressions in
\cite{tosu,aoho,maka},
we can obtain the Hamiltonian of the vibron model from the effective
two--nucleon forces.

Since, however, these relations are rather complicated, the
deduction of the phenomenological parameters from the numerical
values corresponding to specific two--nucleon interactions can
help to illuminate the situation. We have done that for the
microscopic interactions mentioned in Section III.

In order to obtain the phenomenological interaction from the
microscopic one, we can use a set of 4 appropriate matrix
elements, e.g. as shown above; or we can subtract the parameters from
a large number of matrix elements by a fitting procedure.
We have followed the second way, which obviously gives more
reliable parameters. When doing so, we have considered the 101 matrix
elements with the $8 \le n_{\pi} \le 20$ quantum numbers.
{}From such a calculation one obtains in a natural way also a
quantitative measure of the average deviation between the
microscopic and phenomenologic matrix elements.
In Table III we have listed the parameters (in $MeV$) belonging
to the different two--nucleon forces, and in the last column the
average root mean square deviation of the microscopic and
phenomenologic matrix elements are given, too.
Considering the fact that the energy--region of these matrix
elements span more than 100 $MeV$, the deviation is not very
large. In other words, the phenomenological Hamiltonian
containing only first and second order terms can approximate
these effective two--nucleon forces reasonably well.

By comparing these sets of parameters with the one that gave the
best description of the experimental spectrum of Fig. 1. we can
realize some similarities both in the signs and in the overall
magnitudes. But there are some differences as well; the most
remarkable ones are the larger experimental $\beta$ and $\delta$
values, giving rise to more definite splitting with respect to
the $L$ and $n_{\pi}$ quantum numbers.

\section{Summary}

In this paper we have obtained the phenomenological
cluster--cluster interaction of the vibron model from effective
two--nucleon forces. The example we have considered was the
$^{16}O + \alpha$ system, but the same procedure can be applied
to other cases, too. This relation is based on the similarities
between the model spaces of the two descriptions, which was
established by the modification of the basic assumptions of the
vibron model, via selecting out the forbidden states.
The relation can be given in analytical expressions, but they
are too complicated even for the simplest example i.e. for the
case of the
two closed--shell clusters. On the other hand the numerical
values of the phenomenological parameters can be obtained easily
from the effective two--nucleon forces. For the most often used
microscopic interactions we have given the corresponding
Hamiltonians of the vibron model.
\\
\\
This work was supported by the OTKA grant (No. 3010).

\figure{Experimental states of $^{20}$Ne in comparison with the
phenomenological model calculation in terms of $U(3)$ dynamical
symmetry. The dashed lines indicate uncertain band--assignements.}

\begin{table}
\caption{Parameters of the effective nuclear potentials}
\begin{tabular}{ccccccc}
   & $d_k (fm^{-2})$ & $V_k (MeV)$ &  $W_k$ & $M_k$ & $H_k$ & $B_k$
     \\
\tableline
   & 2.5            & -6.0        & 0.4583 & 0.4583 & 0.0417 & 0.0417  \\
HN1& 0.94           & -546.0      & 0.4148 & 0.4148 & 0.0852 & 0.0852  \\
   & 0.54           & 1655.0      & 0.4229 & 0.4229 & 0.0771 & 0.0771  \\
\tableline
   & 2.5            & -6.0        &-0.2361 & 1.1528 & 0.5972 &-0.5139  \\
HN2& 0.94           & -546.0      & 0.4148 & 0.4148 & 0.1310 & 0.0394  \\
   & 0.54           & 1655.0      & 0.4474 & 0.3985 & 0.1015 & 0.0526  \\
\tableline
   & 2.5           & -6.0        &-0.2361 & 1.1528 & 0.5972 &-0.5139  \\
MHN& 0.94           & -546.0      & 0.4240 & 0.4057 & 0.1401 & 0.0302  \\
   & 0.54           & 1655.0      & 0.4474 & 0.3985 & 0.1015 & 0.0526  \\
\tableline
V2 & 1.80           & -60.65      & 0.38   & 0.62   & 0.     & 0.      \\
   & 1.01           &  61.14      & 0.38   & 0.62   & 0.     & 0.      \\
\end{tabular}
\label{table0}
\end{table}

\begin{table}
\caption{SU(3) matrix elements of the Hamiltonian with
Volkov number 2 force for
the $^{20}$Ne (in $MeV$)}
\begin{tabular}{ccccccccc}
$(\lambda,\mu)$      & n=0,   &  1,   &  2,   &  3,   &  4,   &  5,   &  6,   &
 7,  \\
\tableline
(8,0) & 2.97 & 0.33 &-1.90 &-3.46 & -4.16 & --  & --   & --   \\
(9,0) &16.46 &12.93 &10.23 & 8.33 &  7.27 & --  & --   & --   \\
(10,0)&26.75 &23.29 &20.49 &18.41 & 17.08 &16.51& --   & --   \\
(11,0)&37.97 &34.48 &31.44 &29.19 & 27.63 &26.76& --   & --   \\
(12,0)&48.36 &44.80 &41.84 &39.47 & 37.72 &36.61& 36.13 & --   \\
(13,0)&58.70 &55.17 &52.19 &49.74 & 47.85 &46.54& 45.80 & --   \\
(14,0)&68.68 &65.23 &62.26 &59.77 & 57.77 &56.29& 55.34 & 54.94   \\
\end{tabular}
\label{table1}
\end{table}

\begin{table}
\caption{
Parameters of the phenomenological cluster--cluster interactions
obtained from effective two--nucleon forces. The last column
gives the average rms deviation between the microscopic and
phenomenologic matrix elements. (The values are given in $MeV$.)
}
\begin{tabular}{cccccc}
Force&$\beta$&$\gamma$&$\delta$&$\epsilon$&$d$\\
\tableline
V2&0.0493&12.289&-0.119&-91.329&1.15\\
HN1&0.0720&13.776&-0.149&-123.28&1.79\\
HN2&0.0687&13.042&-0.136&-106.01&1.77\\
MHN&0.0562&12.415&-0.121&-93.145&1.46\\
\end{tabular}
\label{table2}
\end{table}
\end{document}